%% file: submit.tex
\documentstyle[pre,aps,preprint]{revtex}
\begin{document}
\draft
\input psfig_19

\title{Phase Transitions in a Forest-Fire Model}
\author{S. Clar, K. Schenk, and F. Schwabl}
\address{Institut f\"ur Theoretische Physik, \\ Physik-Department der
Technischen Universit\"at M\"unchen, \\ James-Franck-Str., D-85747 Garching,
Germany}
\date{\today}
\maketitle
\begin{abstract}
We investigate a forest-fire model with the density of empty
sites as control parameter. The model exhibits three phases,
separated by one first-order phase transition and one 'mixed' phase
transition which shows critical behavior on only one side and hysteresis. 
The critical behavior is found to be that of the self-organized 
critical forest-fire model [B. Drossel and F. Schwabl, Phys. Rev. Lett. {\bf
69}, 1629 (1992)], whereas in the adjacent phase 
one finds the spiral waves of the Bak et al. forest-fire model 
[P. Bak, K. Chen and C. Tang, Phys. Lett. A {\bf 147}, 297 (1990)].
In the third phase one observes clustering of trees with the fire
burning at the edges of the clusters. 
The relation between the density distribution in the spiral state
and the percolation threshold is explained and the implications for 
stationary states with spiral waves in arbitrary excitable systems 
are discussed. Furthermore, we comment on the possibility of mapping
self-organized critical systems onto 'ordinary' critical systems.
\end{abstract}


\narrowtext

\section{Introduction}
 
In 1990, Bak et al. introduced a simple model for the spreading of a fire 
in a forest or the spreading of
disease in a population \cite{bak90}.
The individuals (sites on a square lattice in 2 dimensions) 
can be in one of three states: tree (healthy, excitable), 
tree on fire (infected, excited), and ashes or empty site
(immune or dead, refractory). New individuals are 'fed' into the system
with a small rate $p$. Whether the third state is considered 
as death of an individual and $p$ consequently as the birth rate 
of new individuals, 
or as a state of immunity and $1/p$ as time scale of the loss of that 
immunity, 
is a matter of interpretation.
In the following, we will use the terms 
tree, fire, empty site and refer to $p$ as the tree growth rate.
The exact rules of the model \cite{bak90} 
were as follows: (i) at each time step trees grow at
empty sites with a small probability $p$, (ii) trees on fire will burn down the
next time step and turn to empty sites, 
(iii) the fire on a site will spread to the trees at its
nearest neighbor sites at the next time step.
Although originally claimed to be critical in the
limit $p \to 0$, the simulations in \cite{gra91,mos92} showed that the model 
does not display criticality. Instead, one could observe
quasi-deterministic spiral waves of fires.

In 1992, Drossel and Schwabl introduced the self-organized critical
forest-fire model (SOC FFM) \cite{dro92} 
with the additional rule: (iv) if no nearest
neighbor is burning, a tree catches fire with a small 'lightning' 
probability $f$. Under the condition of a double separation of time scales
(time between two lightning strokes $1/f$ $\gg$ time scale of tree growth 
$1/p$ $\gg$ time needed to burn down large tree 
clusters) the model shows critical behavior over a wide range of 
parameter values. The properties of this model were investigated in e.g.
\cite{hen93,gra93,chr93,cla94}.

In this paper, we shall investigate a model with the same type of interactions
while keeping constant the number of empty 
sites or immune individuals. Their density is the control
parameter of the model. Parameters like $p$ or $f$ do not enter the model.
We will show that the model exhibits three phases,
two of which can be shown to display the same behavior as the abovementioned
models. The third phase shows clustering of trees with the fire
burning at the edges of the clusters.

In the 'spiral wave' phase, 
which exists not only in this model, but in a large 
number of excitable systems (for a review on excitable systems see e.g.
\cite{tys88,mer92}), we will point out an interesting relation to the 
non-equilibrium percolation model of \cite{cla95}. 
There, the tree density was the control parameter and the
following rules were iterated : (i) Lightning strikes an arbitrary site in the
system. If the site is occupied, the whole cluster of $s$ trees which is
connected to this site (by nearest-neighbor coupling) burns down, i.e. the
trees of this cluster turn to empty sites. (ii) Then $s$ new trees are grown at
randomly chosen empty sites (including the ones that have just turned empty).  
The close relation between this model and the model treated in the present 
paper will enable 
us to identify the mechanism that determines the density distribution of
excitable constituents and, in particular, the density
immediately in front of excitation fronts in the spiral-wave phase of 
excitable systems.

Furthermore, the model presented in this paper, as well as the model in
\cite{cla95}, are examples of reformulations of a SOC model
in terms of a control parameter (the density of trees or empty sites,
respectively). Both models indicate that the claim in
\cite{sor95} that all SOC models can be mapped onto ordinary critical
systems exhibiting a sub-critical phase, a critical phase with a smoothly
varying order parameter and a critical point that separates
the two phases, is not true in general. Non-equilibrium systems and
their phase transitions show a much richer behavior than equilibrium
systems, with many features that are unknown in equilibrium.

\section{The model}

The model is defined on a $d$-dimensional hyper-cubic lattice with $L^d$
sites. If not stated otherwise, we choose $d=2$ and periodic boundary 
conditions in the following. 
$\rho_e L^d$ sites are randomly chosen to be empty. The density of empty sites
$\rho_e$ is the control parameter
of the model. The remaining sites are randomly filled with trees and fires. 
Their densities are denoted by $\rho_t$ and $\rho_f$.
It is always $\rho_t + \rho_e + \rho_f = 1$. 
The exact values of $\rho_t$ and $\rho_f$ in the initial state do not
affect the stationary state, except in the vicinity of
certain points which will be discussed later.

The system is iterated as follows: (i) all trees on fire will burn down the
next time step, (ii) the fire on a site will spread to the trees at its
nearest neighbor sites in the next time step, (iii) after each time step
the same number of trees that have burnt down grow at randomly 
chosen empty sites (including the ones which have just become empty), 
thereby keeping $\rho_e$ fixed, and (iv) if the fire
dies out, a randomly chosen tree catches fire spontaneously. 

The motivation of rule (iv) is the following:
We want to investigate the system 
under the action of a vanishingly small lightning probability .
Since the process which is described by rule (iv) may then take very long in 
real time, but nevertheless can be simulated in one iteration step,
rule (iv) represents an acceleration of the real process. When calculating
temporal averages of the fire density, this point has to be considered.

The reason for choosing the density of empty sites as parameter and not
the density of trees is the following: Consider a system that consists
only of trees and replace one tree by a fire. In the next step we have four 
fires, but only one empty site to grow new trees. This extreme example
shows that there might be situations in which it is not possible
to keep the density of trees constant.
The density of empty sites, in turn, can always be kept constant for 
arbitrary values within the interval $[0,1]$. 

In the following sections, we discuss the properties of the stationary 
state as function of the density of empty sites $\rho_e$. 
In Sec.~\ref{secthree}, we start with a high density of empty sites
and investigate the region of vanishing fire density. In Sec.~\ref{secfour},
we lower the density of empty sites which leads to a state with spiral waves.
A detailed description of the mechanism which determines the density
distribution in the spiral state is given in Sec.~\ref{secfive}. 
Sec.~\ref{secsix} investigates the more homogeneous, 'mixed' phase that can 
be observed 
after further decreasing the density of empty sites. Up to here, only
two-dimensional square lattices are considered. 
Sec.~\ref{secseven} treats other dimensions and lattice types. In
Sec.~\ref{seceight}, we comment on the issue of mapping self-organized
criticality onto ordinary criticality. Finally, in the appendix, we 
present some general properties of the order parameter curve. 

\section{Region of Vanishing Fire Density and Critical Point}
\label{secthree}

For $\rho_e \lesssim 1$,
there exist only very small tree clusters, and, consequently, if one starts 
a fire
by setting on fire a randomly chosen tree, it soon dies out, and one has to
start a new one. The average number of trees $\bar s$ destroyed by a fire 
therefore
is finite and small, and in the thermodynamic limit $L \to \infty$ the fire 
density $\rho_f$ equals zero
(taking into account that rule (iv) is an acceleration of a process which
takes infinitely long in real time). 
In our simulations the maximum system size was 
$6000^2$. 
With decreasing $\rho_e$, $\bar s$ increases, but still remains finite, and
therefore $\rho_f = 0$. If $\rho_e$ is decreased further, we finally arrive 
at a critical density $\rho_e^{c,1} \approx 59.2\%$ 
($\rho_t^{c,1} = 1 - \rho_e^{c,1} \approx
40.8\%$), where $\bar s$ diverges with a power law 
$\bar s \propto (\rho_e - \rho_e^{c,1})^{-\delta}$, with some exponent 
$\delta$. 
A snapshot of the system in the vicinity 
of $\rho_e^{c,1}$ is shown in Fig.~\ref{luisa}.

The critical behavior close to $\rho_e^{c,1}$ 
can be described by exponents which
are defined as in percolation theory \cite{sta92}. The size distribution of
tree clusters is $n(s) \propto s^{-\tau} {\cal C}(s / s_{\text{max}})$ with
a cutoff function ${\cal C}$. $s_{\text{max}}$ is the size of the
largest cluster in the system and diverges for $\rho_e \to \rho_e^{c,1}$. 
The fractal dimension $\mu$ 
of the clusters is defined by $R(s) \propto s^{1/\mu}$,
where $R$ is the radius of gyration of a cluster.
The correlation length is given by $\xi \propto (\rho_e - 
\rho_e^{c,1})^{-\nu}$.
More exponents can be defined and scaling relations between them can be 
derived (see e.g. \cite{hen93,gra93,chr93,cla94}).

The critical exponents found in the simulations ($\tau = 2.14(4)$, 
$\mu = 1.95(2)$, and $\nu = 0.28$) are the same as in the SOC FFM, when 
appropriately redefined (for $\nu$
one has to change variables from $f/p$ to $\rho_e$ via 
$\rho_e - \rho_e^{c,1} \propto (f/p)^{1/\delta}$ (see \cite{cla94})). 
Also the critical density $\rho_t^{c,1} \approx 40.8\%$ remains the same.
This model displays exactly the same critical behavior as the SOC FFM and
the model of \cite{cla95}, because
in a system which is much larger than the correlation length, 
neither the difference
between a globally conserved density $\rho_e$ (this model and \cite{cla95}) 
and a density $\rho_e$ 
which is only conserved on an average (SOC FFM) 
nor the difference between instantaneous re-growth of trees (this model) or 
delayed re-growth (SOC FFM and \cite{cla95}) can be seen on length scales 
comparable to the correlation length. 

In the stationary state of the SOC FFM $\bar s = p \rho_e / f (1 - \rho_e)$, 
since in one time step there are $\rho_t L^d f$ lightning strokes
and $\rho_e L^d p$ new trees are growing. 
Therefore, if we measure in our model $\bar s$ 
for a certain value of the control
parameter $\rho_e$ we know that its behavior is 
that of the SOC FFM for $f/p = \rho_e/ \bar s (1 - \rho_e)$. 

\section{Region of Finite Fire Density I: Spirals}
\label{secfour}

\begin{figure}
\psfig{file=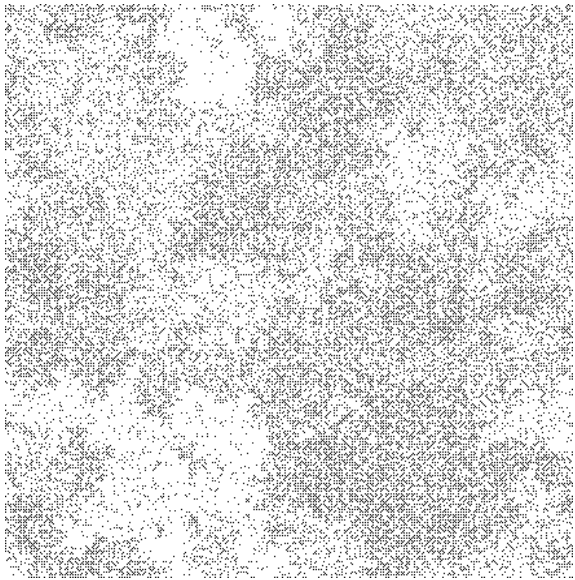,height=3in,angle=0}
\vskip 1cm
\caption{Snapshot of the stationary state in the 'SOC' phase 
near the critical density $\rho_e^{c,1} \approx 59.2\%$. 
$L = 2000$ and $\rho_e = 59.8\%$. 
Trees are black and empty sites are white.}
\label{luisa}
\end{figure}

\begin{figure}
\psfig{file=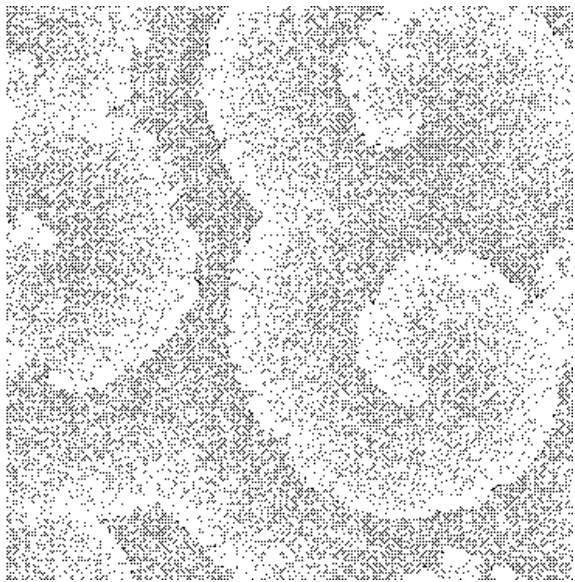,height=3in,angle=0}
\vskip 1cm
\caption{Snapshot of the stationary state in the 'spiral wave' phase 
for
$L = 1000$ and $\rho_e = 59\%$. 
Trees are grey and empty sites are white. The fires are black, but difficult to
see. They are located at those lines where the density of trees changes 
abruptly.} 
\label{carmen}
\end{figure}

If we decrease the number of empty sites beyond the critical point 
$\rho_e^{c,1}$, the size of the largest forest cluster diverges, and we would
expect the fire not to be extinguished any more. Therefore, 
one might expect the fire density
$\rho_f$ to behave as an order parameter that 
sets in at $\rho_e^{c,1}$ and grows smoothly from zero 
to finite values, obeying some power law $\rho_f \propto 
(\rho_e^{c,1} - \rho_e)^\beta$.
Instead, the following behavior can be observed in the simulations: 
The system restructures, fires gathering
to form spirals and the regions of different densities 
(see Fig.~\ref{luisa}) vanishing in favor of a smooth density distribution
between the spiral arms (see Fig.~\ref{carmen}). 
The behavior of the spirals is quasi-deterministic.
Immediately in front of the
fire fronts the tree density is very 
high and immediately behind them it is very low. 
The density distribution is treated in more detail in Sec.~\ref{secfive}.
The distance $\Delta$ between two spiral arms is finite and 
constant throughout the system. The fire density
for $\rho_e = \rho_e^{c,1} - 0$ is also finite.  
Decreasing $\rho_e$ further, the distance
between the spiral arms becomes smaller and the fire density 
becomes larger, since it is now easier for the fire to survive. 
At $\rho_e = \rho_e^{c,3} \approx 54.2\%$, 
the spiral state breaks down and
another restructuring to a new phase takes place, which will be treated in
Sec.~\ref{secsix}.

For $\rho_e \le \rho_e^{c,1}$ the fire is able to sustain itself and
it is no longer necessary to keep it alive by setting on fire randomly chosen
trees each time the fire has died out. The 'external field' can be set zero.
In terms of tree growth and lightning, 
we have a model with tree growth rate $p$, but without lightning rate $f$.
Therefore, the behavior in the spiral state at some density $\rho_e$ 
is exactly the same as in the 
forest-fire model of Bak et al. in \cite{bak90} for a certain
value of the tree growth probability $p$, since
in the thermodynamic limit it does not matter which of the variables is
kept fixed. The fluctuations of the densities in the Bak model vanish for
$L \to \infty$, as do the fluctuations of the number of new grown trees
per time step in this model.
 
In the Bak model, one observes a diverging distance between 
spiral arms for $p \to 0$. If we start in our model 
with a spiral state and increase 
$\rho_e$ beyond $\rho_e^{c,1}$ (corresponding to a decrease of $p$), 
the system
does not undergo a phase transition at $\rho_e = \rho_e^{c,1}$, 
but chooses to retain the spiral state
up to $\rho_e = \rho_e^{c,2} \approx 60.8\%$. 
This value of $\rho_e$ with $\Delta = \infty$ 
corresponds to $p = 0$ in the Bak model.
As function of $p$ the order parameter follows
the power law $\rho_f \propto p$ (explanation in Sec.~\ref{secfive}).

As already found in \cite{gra91,mos92}, 
the spiral state does not display criticality or scale
invariance in the sense of clusters or events on all length scales. 
For a particular value of $p$ the model does not show events on all 
scales, but only on the scale $1/p$, and behaves essentially
deterministic.
However, the model allows similarity transformations, since a system with 
parameter $p_1$ can be obtained by rescaling a system with parameter $p_2$. 

\begin{figure}
\psfig{file=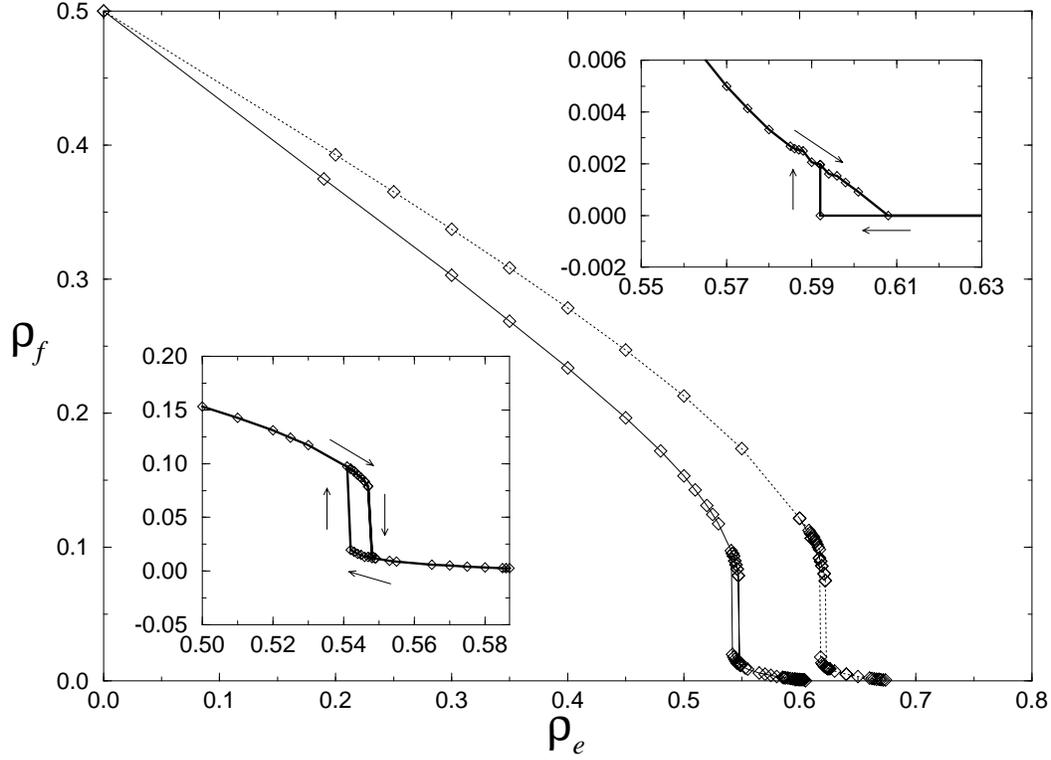,height=4in,angle=-90}
\vskip 1cm
\caption{The order parameter fire density $\rho_f$ as function of the density 
of empty sites $\rho_e$ in two dimensions (---, square lattice; $\cdots$, 
triangular lattice). The magnified sections in the insets 
show the two transitions of the
two-dimensional system on a square lattice. To the left, one can see the
first order phase transition at $\rho_e^{c,3}$ and $\rho_e^{c,4}$ between the
mixed phase and the spiral phase, and to the 
right the 'mixed' phase transition at $\rho_e^{c,1}$ and $\rho_e^{c,2}$ between
the spiral phase and the SOC phase. 
The arrows indicate the directions in which the transitions are traversed.}
\label{2dphasediagram}
\end{figure}

What we have found is a discontinuous
phase transition which is neither of first nor of
second order. We find hysteresis (the state of the system for a density
in the interval $[\rho_e^{c,1};\rho_e^{c,2}]$ depends on from which side one 
approaches this region), an order parameter that increases smoothly from zero 
at $\rho_e^{c,2}$ to finite values, and critical behavior 
on only one side of the transition 
($\rho_e \ge \rho_e^{c,1}$, the side with vanishing order parameter). 
The order parameter curve for a two-dimensional system 
is shown in Fig.~\ref{2dphasediagram}.

The reason why the critical point $\rho_e = \rho_e^{c,1}$ is not an ordinary 
critical point is the same as in \cite{cla95}. From Fig.~\ref{luisa}, 
it can be seen that, in addition to large tree clusters, there exist also 
large clusters of empty sites. In contrast to ordinary critical phenomena 
(e.g. percolation), there is no homogeneously distributed set of large 
clusters that could join at $\rho_e = \rho_e^{c,1}$ to form an infinite 
cluster that spans the whole system. Rather, 
the largest cluster has to compete with all other regions
with different densities (which, due to the very nature of the dynamics, 
are nothing more than different
growth stages of the largest cluster itself) for space in the system.
Or, to put it differently, an infinite cluster like in
percolation is impossible, because the system has to provide
space not only for the infinite cluster, but also for a large
number of 'younger' copies of it. 
The violation of the hyperscaling relation $d = \mu (\tau - 1)$
also indicates the inhomogeneous distribution of density in the
forest-fire model. As explained in \cite{hen93}, 
the violation is equivalent
to the statement that not every part of the system contains a spanning 
cluster at criticality. 

\section{Density Distribution in the Spiral State}
\label{secfive} 

\begin{figure}
\psfig{file=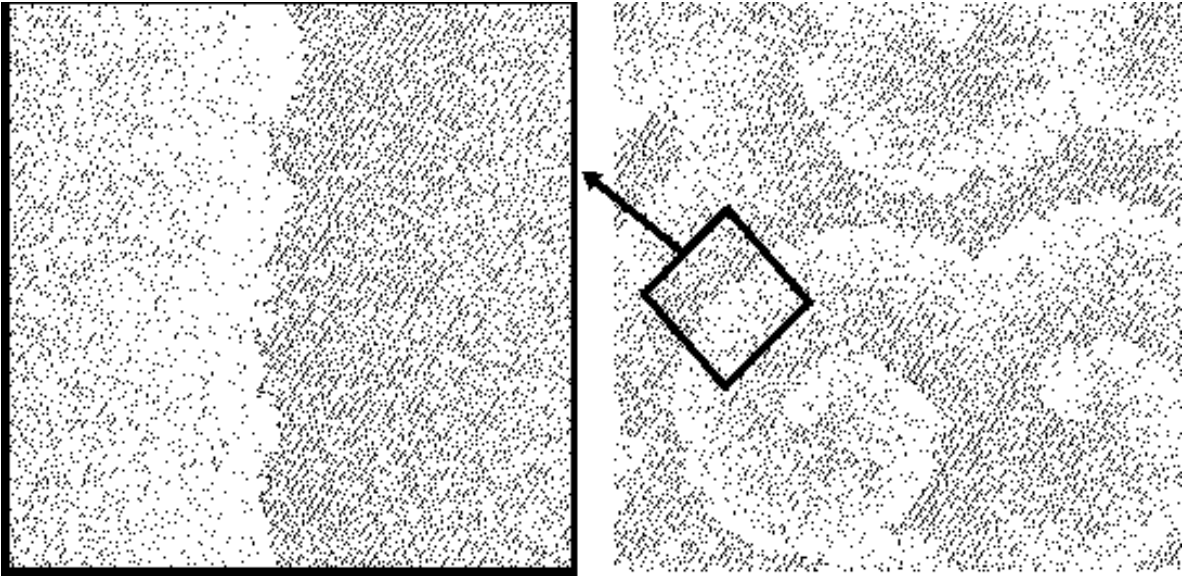,height=3in,angle=-90}
\vskip 1cm
\caption{A spiral state at density $\rho_e = 59\%$ with distance $\Delta$ 
between successive fire fronts can be covered by quadratic 'single front' 
sections of edge length $\Delta$. 
Trees are grey, fires are black, and empty sites are white.}
\label{arantza}
\end{figure}

In order to understand better the density 
distribution in the spiral state, we shall now investigate a 
state with a single front propagating through a square system
with periodic boundary conditions (see left side of Fig.~\ref{arantza}).
Since this can be considered to be a section of a spiral state
(see right side of Fig.~\ref{arantza}), and the
spiral state, in turn, can be completely covered by such sections (if the
edge length is chosen to be equal to the distance $\Delta$ 
between two successive 
arms of the spirals), it is sufficient to understand the density
distribution in this part of the whole system.
Instead of successive fire fronts passing through the small section of
the system, one might also think of a single fire front which repeatedly
leaves the section at one end and reenters at the other end due to
periodic boundary conditions.
 
The tree density immediately in front of the fire front 
$\rho_t^{\text{before}}$ 
has to be larger than or equal to the percolation threshold for site
percolation on a square lattice $p_c \approx 0.59$, 
for otherwise there wouldn't be 
propagation. With increasing distance from the fire front,
the tree density smoothly decreases, until one finally arrives again on
the other side of the front, where the density takes its lowest
value $\rho_t^{\text{after}}$. If $\rho_t^{\text{before}}$ was exactly the 
percolation threshold,
the fire would burn down a vanishing fraction of all trees, 
and the propagation speed of the front 
would be zero. The densities in front of and behind the fire front would
be equal ($\rho_t^{\text{before}} = p_c = \rho_t^{\text{after}}$), 
as claimed in \cite{gra91}. 
However, the simulations clearly show that this is not the case.
$\rho_t^{\text{before}}$ is higher than the percolation threshold, 
the density $\rho_t^{\text{after}}$ in a region
after the fire has passed through it is very low, but not equal to zero, 
and the propagation
speed of the fire fronts $v_{\text{fire}}$ is non-vanishing ($0.7 \pm 0.05$
sites per iteration step near $\rho_e^{c,2}$). 
What is the mechanism that determines the density distribution and 
the values of the density immediately in front of and behind the fire front? 

We shall approximate the periodic 'single front' state even further 
by a coarsened state with a 
finite but large number $n$ of stripes of equal width and different densities
parallel to the fire front. Each stripe has homogeneous density.
Let $\rho_t^1, \ldots, \rho_t^n$ be the densities of the $n$ stripes, 
starting with the highest density. $\rho_t^i$ is then the average density 
of the original smooth 'single front' system in the region covered by 
the $i$th stripe ($\rho_t^1 \approx \rho_t^{\text{before}}$ 
and $\rho_t^n \approx 
\rho_t^{\text{after}}$). We also coarsen time, in that we consider the 
propagation
of the fire front from the bottom of one stripe to its top as one 
time step. Growing of new trees then takes place after each
(coarsened) time step. 
If one sets on fire the baseline of the stripe with highest density,
the infinite cluster in that stripe will surely be set on fire, 
since it has many
connections with the baseline, and the propagation of the fire front 
effectively causes the removal of the infinite cluster from this stripe. 

The propagation of the fire front therefore can be modeled as follows. 
First, we identify the infinite cluster in the stripe with highest density
and remove it from the system. Doing this one has to respect the
boundaries of the stripes, although the infinite cluster of course
extends into the neighboring stripes. The strength of the infinite cluster
at density $\rho_t^{1}$ is denoted $P(\rho_t^1)$.
(We now define also the density 
$\rho_t^{n+1}$ of the stripe which contained the infinite cluster after
its removal, i.e. $\rho_t^{n+1} = \rho_t^{1} - P(\rho_t^1)$.) 
Second, the $P(\rho_t^1) \Delta^2/n$ trees of the infinite cluster
are redistributed randomly amongst the empty sites of the
whole system of size $\Delta^2$.  
If the system is to be stationary, the stripes thereby just 
exchange their densities, i.e. the stripe with second highest 
density $\rho_t^2$ now assumes the highest density $\rho_t^1$, 
and so on. From that condition, one can easily derive the equation
\[
\rho_t^{i-1} = \rho_t^i + (\rho_t^1 - \rho_t^{n + 1}) \cdot 
(1 - \rho_t^i)/(n (1 - \rho_t) + \rho_t^1 - \rho_t^{n + 1})
\]
for $i = 2, \ldots, n + 1$.
The last factor on the r.h.s. represents the fraction of trees of the 
infinite cluster that are re-grown in the stripe with density $\rho_t^i$.
We finally obtain
\begin{equation}
\frac{1 - \rho_t^1}{1 - \rho_t^2} = \frac{1 - \rho_t^2}{1 - \rho_t^3} 
= \ldots =
\frac{1 - \rho_t^n}{1 - \rho_t^{n+1}}. \label{siegfried}
\end{equation}
Together with $\rho_t = (1/n) \sum_{i=1}^n \rho_t^i$ 
we have $n$ equations for $n+1$ densities. 
These equations were already derived in \cite{cla95} for a related model.
The average density in a system with $n$ stripes is then:
\begin{eqnarray*}
\rho_t &=& 1 - \frac{1 - \rho_t^{n+1}}{n} \sum_{i=1}^{n} 
{\left(\frac{1 - \rho_t^1}{1 - \rho_t^{n+1}}\right)}^{i/n} \\
&=& 1 - \frac{\rho_t^1 - \rho_t^{n+1}}{n(((1-\rho_t^1)/
(1-\rho_t^{n+1}))^{-1/n}-1)}
\end{eqnarray*}
In the system that we are really interested in, the density varies smoothly, so
we have to consider large $n$.
In \cite{cla95} it was argued that $\rho_t^1$ has always to be greater 
than or equal to a certain constant $\rho_t^* \approx 0.625 > p_c$. If this
were not the case, a traversing fire front would leave behind large tree 
clusters which would lead to inhomogeneities that prevented the next
fire front from passing through in the same quasi-deterministic way than
the first one.  
Likewise, $\rho_t^{n+1}$ has always to be greater than or equal 
to another constant $\rho_t^{\infty} = \rho_t^* - P(\rho_t^*) \approx 0.078$.
$\rho_t^*$ is always larger than
the corresponding percolation threshold $p_c$. 
These constants
have a more fundamental significance independently of this model
and of any states with spiral waves (for details see \cite{cla95,cla96}). 
The case $\rho_t^1 = \rho_t^*$ and $\rho_t^{n+1} = \rho_t^{\infty}$ 
corresponds to the lowest possible 
overall density and therefore to the state with infinite spirals at $p = 0$
or $\rho_e = \rho_e^{c,2}$.
In this case, we have $\rho_t^{\text{before}} = \rho_t^*$ and
$\rho_t^{\text{after}} = \rho_t^{\infty}$.


The overall density in the spiral state can then be written as
\begin{eqnarray}
\rho_t &=& 
1 - \frac{\rho_t^{\text{before}} - \rho_t^{\text{after}}}{\lim_{n \to \infty} 
n ((\frac{1-\rho_t^{\text{before}}}{1-\rho_t^{\text{after}}})^{-1/n} - 1)} 
\nonumber \\
&=& 1 - \frac{\rho_t^{\text{before}} - \rho_t^{\text{after}}}{\lim_{n \to 
\infty} n (1 - \ln(\frac{1-\rho_t^{\text{before}}}{1-\rho_t^{\text{after}}})/n
\pm \ldots - 1)} \label{soledad} \\
&=& 1 - \frac{\rho^{\text{before}} - \rho_t^{\text{after}}}{\ln(\frac{1-
\rho_t^{\text{after}}}{1- \rho_t^{\text{before}}})} \nonumber 
\end{eqnarray}
For $\rho_t^{\text{before}} = \rho_t^*$ and 
$\rho_t^{\text{after}} = \rho_t^{\infty}$,
this density should be equal to the density $\rho_t^{c,2}$ of the state with
spirals of infinite extension.
With the values of $\rho_t^*$ and $\rho_t^\infty$ measured in \cite{cla95}
($0.625$ and $0.078$ for the square lattice, $0.533$ and $0.062$ for the
triangular lattice), 
one arrives at 
$\rho_t^{c,2,\text{calculated}} = 0.392 \pm 0.002$ for the square lattice
and $\rho_t^{c,2,\text{calculated}} = 
0.325 \pm 0.002$ for the triangular lattice, 
in excellent agreement with the values 
$\rho_t^{c,2} = 0.392 \pm 0.002$ for the square lattice 
and $\rho_t^{c,2} = 0.323 \pm 0.005$ for the triangular lattice,
measured in the spiral state of this model.  

Since Eq.~(\ref{siegfried}) represents a geometric series, the density as 
function of the distance from the fire front is given by
\begin{equation}
\rho_t(x) = 1 - (1 - \rho_t^{\text{after}}) 
\left(\frac{1 - \rho_t^{\text{before}}}{1 - 
\rho_t^{\text{after}}}\right)^{x / \Delta},
\label{eva}
\end{equation}
with $\rho_t(0) = \rho_t^{\text{after}} \gtrsim \rho_t^\infty$ 
and $\rho_t(\Delta) = 
\rho_t^{\text{before}} \gtrsim \rho_t^*$.

With our new knowledge about the nature of the spiral state, we can
derive an equation for the distance $\Delta$ between the fire fronts
as function of the tree growth probability $p$. 
Let the speed of the fire fronts $v_{\text{fire}}(\rho_t^{\text{before}})$ 
be measured in sites per iteration step. Like $P$ it depends on the density
in front of the fire front $\rho_t^{\text{before}}$.
The amount of matter burnt in unit time is then $P(\rho_t^{\text{before}})  
v_{\text{fire}}(\rho_t^{\text{before}}) \Delta$.
This has to be equal to the number of growing trees per unit time
$(1 - \rho_t) p \Delta^2$, leading to the relationship
\begin{equation}
\Delta = \frac{P(\rho_t^{\text{before}}) 
v_{\text{fire}}(\rho_t^{\text{before}})}{1 - \rho_t} \frac{1}{p}.
\label{guadalupe}
\end{equation}
This confirms the observed scaling behavior $\Delta \propto p^{-1}$
and additionally delivers the constant of proportionality, being
$0.63 \pm 0.05$ 
%
%
for $\rho_t^{\text{before}} = \rho_t^*$ 
and in good agreement with simulations.
Since the fire density is inversely proportional to the distance between the
spiral arms, we find $\rho_f \propto p$, i.e. the order parameter exponent
$\beta$ equals one. This can also be seen from the equality 
$\rho_f = p \rho_e$ \cite{dro93}, which states that the number of burnt trees
has to be equal to the number of new grown trees in the stationary state.

If one chooses a fixed site behind the front and wants to know the tree
density in this region as function of the time $t$ until at $T = \Delta
/ v_{\text{fire}}$ the next front passes through, one has to replace in 
Eq.~(\ref{eva}) $x$ by $v_{\text{fire}} t$ and arrives with 
Eq.~(\ref{guadalupe}) at 
\begin{displaymath}
\rho_t(t) = 1 - (1 - \rho_t^{\text{after}}) 
\left[\left(\frac{1 - \rho_t^{\text{before}}}{1 - 
\rho_t^{\text{after}}}\right)^{\frac{1 - 
\rho_t}{P(\rho_t^{\text{before}})}}\right]^{p t},
\end{displaymath}
which with Eq.~(\ref{soledad}) and $P(\rho_t^{\text{before}}) =
\rho_t^{\text{before}} - \rho_t^{\text{after}}$ leads to
\begin{equation}
\label{angela}
\rho_t(t) = 1 - (1 - \rho_t^{\text{after}}) e^{-p t}. 
\end{equation}
Eq.~(\ref{angela}) can also be derived easily from 
$\partial \rho_t / \partial t = p (1 - \rho_t)$, 
first stated in \cite{gra91}. This shows that our picture is in accordance
with the basic equations describing the spiral state.

One additional point concerning Eq.~(\ref{guadalupe}) should be mentioned.
Eq.~(\ref{guadalupe}) relates $\Delta$, $\rho_t$, and $p$. If one regards
e.g. $\Delta$ as fixed, only the product $(1 - \rho_t) p$ is determined.
This reflects the fact that in a 'single front' state with fixed edge length
$\Delta$ one can have different overall densities $\rho_t$. The other 
quantities adjust themselves automatically. However, in the 'real' spiral 
state, fixing $\Delta$ determines invariably all other quantities. Since
the spiral state can be constructed by connecting 'single front' systems, this
seems to be a contradiction. The solution of this apparent contradiction is
that one has neglected the spiral centers. They yield a second, unfortunately
unknown, relation between $\Delta$, $\rho_t$, and $p$. This relation is brought
about by the rotation of the spiral centers. $\rho_t$ and $p$ determine the
angular velocity $\omega$ of the rotation, which, in turn, determines $\Delta$
via $\Delta = v_{\text{fire}} T = v_{\text{fire}} 2 \pi / \omega$. For the
calculation of $\rho_t^{c,2}$ with Eq.~(\ref{soledad}), using the decomposition
into 'single front' states, it is justified to neglect the spiral centers,
because in the thermodynamic limit their number density is zero. Nevertheless,
one has to be aware of the fact that, although not important for calculating 
$\rho_t^{c,2}$, the spiral centers are the 'pacemakers' of the spirals and
therefore responsible for the magnitude of $\Delta$. 

Another interesting point is that the density $\rho_e^{c,2}$ can also be
found from an extremum principle first stated in \cite{dro92}. 
There, the extremum
principle was erroneously used to determine the critical density 
$\rho_t^{c,1}$ of the SOC FFM. 
The fire was believed to destroy as much trees as it can at the 
critical point, but the result $\approx 39\%$ was in contradiction to
the measured value $\rho_t^{c,1} \approx 40.8\%$ 
\cite{hen93,gra93,chr93,cla94}. 
However, the equations derived in \cite{dro92} from the extremum principle
can easily be seen to be equivalent to Eq.~(\ref{soledad}) 
for the spiral state. Therefore, the 
principle yields the correct critical density, 
but for a different, then unknown, phase of the model. 

From the results found in this section one can draw some conclusions for  
excitable media in general. In many excitable systems
stationary states with spiral waves can be found.
Famous examples are e.g. the Belousov-Zabhotinski (BZ) reaction \cite{ros88} or
the electro-physiological activity of heart tissue \cite{dav92}.
If the spirals are to sustain themselves in a stationary state, the
density of the excitable constituents (the 'fuel') 
in a region immediately before a
fire front passes through it $\rho_t^{\text{before}}$ 
has always to be larger than or equal to 
some threshold $\rho_t^*$, which, in turn,
is larger than the percolation threshold $p_c$ 
for that particular situation. The percolation threshold can in principle
be measured by preparing a homogeneous system with a certain
density of excitable constituents and 'exciting' one edge of the system.
If the resulting excitation front dies out before reaching the other end
of the system, we are still below the percolation threshold, and if it 
reaches the other end in finite time we are above. At the percolation
threshold the front barely survives and needs an infinitely long time to
reach the other end. The overall density of the excitable constituents
has to be above the value $\rho_t^{c,2}$, below which no excitation
can be sustained. 

A system that displays spiral waves or avalanches (concentric waves, target
patterns) depending on the density of excitable constituents 
are e.g. populations of Dictyostelium Discoideum
amoebae \cite{lee96}. There, circular waves of signaling activity
emanating from pacemakers
are found for low cell densities, whereas for high cell densities one observes 
spiral waves. 

\section{Region of Finite Fire Density II: The Mixed Phase}
\label{secsix}

\begin{figure}
\psfig{file=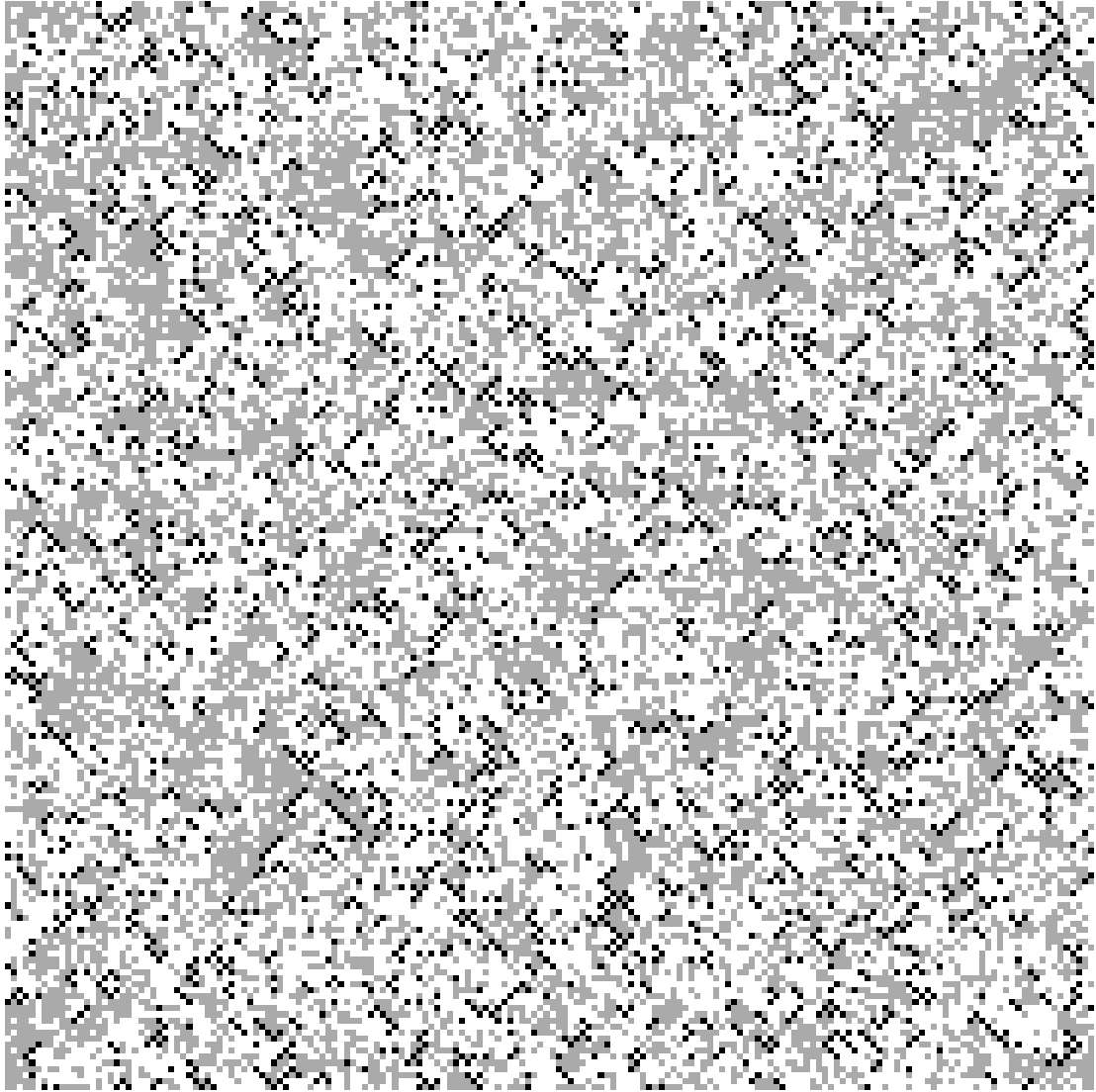,height=6in,angle=0}
\vskip 1cm
\caption{Snapshot of the stationary state in the 'mixed' phase
near the critical density $\rho_e^{c,4} \approx 54.7\%$. 
$L = 180$ and $\rho_e = 54.6\%$. 
Trees are grey, fires are black, and empty sites are white.}
\label{covadonga}
\end{figure}

With decreasing $\rho_e$, the distance between the spiral arms becomes 
smaller and the spirals finally break
up to single fronts. This change occurs continuously. The fronts, which
are more irregular than the spirals, have also been observed in the model of
Bak et al. Decreasing $\rho_e$ further, the coherence length of the fronts
soon becomes comparable with the lattice constant, and the 
system reorganizes itself discontinuously
into another state, which constitutes the third phase of this model.
This happens at $\rho_e = \rho_e^{c,3} \approx 54.2\%$.
The simulations show that the trees in the new state tend to 'cluster' 
(to form regions with higher tree density) with the
fire burning at their edges (see Fig.~\ref{covadonga}). 
To sustain this type of structure, a certain minimum density of fires is
needed. 
At $\rho_e = \rho_e^{c,3}$, $\rho_f$ 
jumps from $2\%$ to $10\%$. 
If one lowers $\rho_e$ further, the size of the 'clusters' decreases, and
the fire density increases. The system as a whole becomes more homogeneous
(trees, fires and empty sites are 'mixed').
For $\rho_e = 0$, we observe $\rho_t = \rho_f = 1/2$ for all dimensions and
lattice types. The order parameter curve starts 
linearly at $\rho_e = 0$ 
with a slope that depends only on the number of nearest neighbors $z$
(explanations in the appendix). 
If we start with small $\rho_e$ and
traverse the phase transition in the other direction, it takes
place at different $\rho_e$ ($\rho_e^{c,4} \approx 54.7 \% > \rho_e^{c,3}$), 
i.e. one has a first-order phase transition with hysteresis.
The order parameter as function of the density of empty sites 
is shown in Fig.~\ref{2dphasediagram}.

\section{Dimensions other than Two and Different Lattice Types}
\label{secseven}

\begin{figure}
\psfig{file=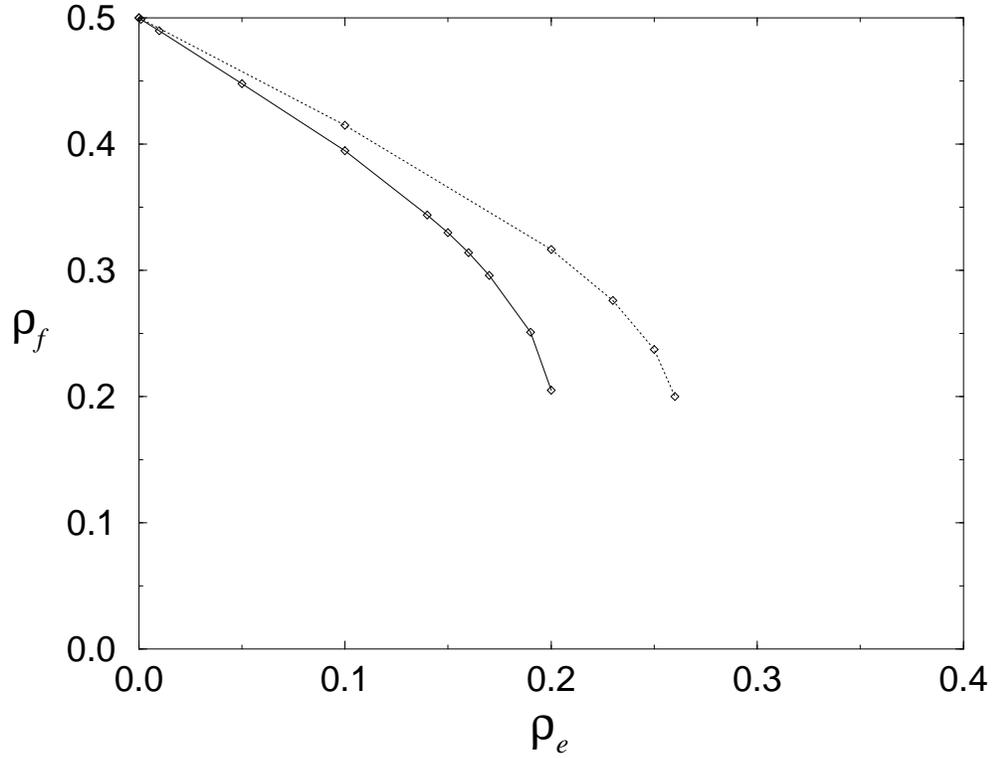,height=4in,angle=-90}
\vskip 1cm
\caption{The order parameter fire density $\rho_f$ as function of the density 
of empty sites $\rho_e$ in one dimension (---, nearest neighbor interaction;
$\cdots$, the fire is allowed to jump over one empty site if necessary).
For $\rho_e > \rho_e^{c,4} \approx 0.2$ and $\rho_e > \rho_e^{c,4} \approx 
0.26$, respectively, the fire dies out.
In the reverse direction, $\rho_f$ remains
zero until $\rho_e = \rho_e^{c,1} = 0$.}
\label{1dphasediagram}
\end{figure}

Since in one dimension there cannot exist spiral waves, we expect a simpler
phase diagram than in two dimensions. 
For $\rho_e = 0$ we are dealing with a completely deterministic one-dimensional
cellular automaton where each site can be in one of two states (automaton 
no. 54, according to the classification scheme of Wolfram \cite{wol83}). 
In the stationary state each tree has at least one fire as neighbor and
vice versa. Strings of more than two trees or fires are not stable, since
sooner or later they would be invaded by fires. The
stationary state is periodic with period two.
The introduction of empty sites makes the automaton non-deterministic, because
the trees can now choose where to grow. This variant has not been investigated
so far.
In the simulations, system sizes of up to $L = 10^7$ were used.
The measured order parameter curves for the one-dimensional case 
can be seen in Fig.~\ref{1dphasediagram}. In addition to the usual 
nearest-neighbor interaction, we simulated also a variant where the fire
is allowed to jump over one empty site if necessary.
The curves start linearly 
at $\rho_f = \rho_t = 1/2$ with different
slopes (explanation see Appendix). 
For increasing $\rho_e$ one observes the same phenomena as in two
dimensions. The tree clusters (strings in 1D) become larger and are 
accompanied by fire on at least one of their ends. At $\rho_e = \rho_e^{c,4} 
\approx 20\%$ 
($\rho_e^{c,4} \approx 26\%$ for next-nearest neighbor interaction) the density
of empty sites is too high for this structure to survive, and the fire 
density drops from a finite value of approximately $20\%$ to zero, since the
system does not have the possibility to rearrange itself into 
a spiral wave phase. Instead, now isolated chunks of trees are ignited
and burn down in finite time. In the reverse direction, $\rho_f$ remains
zero until $\rho_e = \rho_e^{c,1} = 0$, 
i.e. the hysteresis in one dimension is maximum.
In the vicinity of the point $(\rho_e = 0, \rho_f = 0)$, the critical 
behavior of the SOC FFM in one dimension \cite{dro93a} is reproduced.
Since there exists no
spiral phase in one dimension, there are no $\rho_e^{c,2}$ and $\rho_e^{c,3}$.
                          
The two-dimensional simulations were also done with a triangular lattice and 
yielded the same behavior as for the square lattice (for the order parameter
curve see Fig.~\ref{2dphasediagram}). 
The values of $\rho_e^{c,1,2,3,4}$ and
the slope of the order parameter curve for $\rho_e \to 0$ of course
are different (see Tab.~\ref{table1}).

\begin{figure}
\psfig{file=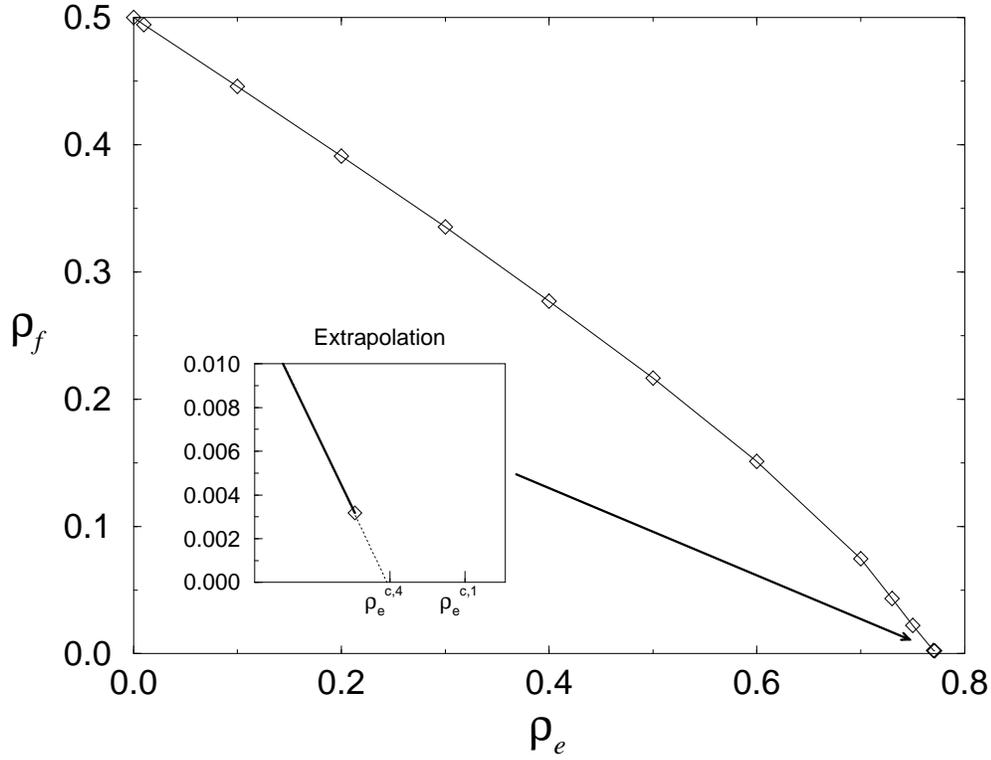,height=4in,angle=-90}
\vskip 1cm
\caption{The order parameter fire density $\rho_f$ as function of the density 
of empty sites $\rho_e$ in three dimensions. The extrapolation of the fire
curve yields the critical density $\rho_e^{c,4} \approx 77.1\% \pm 0.2\%$.} 
\label{3dphasediagram}
\end{figure}

In three dimensions, we could identify a sub-critical phase 
as well as a mixed phase. The order parameter curve is plotted in 
Fig.~\ref{3dphasediagram}. 
Due to small $L$ ($L \le 300$) the critical exponents in the subcritical phase
and the critical density $\rho_e^{c,1}$ could not be measured with 
sufficient accuracy.
The mixed phase in three dimensions shows analogous behavior to the
two-dimensional mixed phase and does not display new phenomena.
The fire density curve could not be measured for very small $\rho_f$
($\rho_f \lesssim 0.2\%$), due to finite size effects.
If one extrapolates the fire density curve to the point $\rho_f = 0$, one
arrives at $77.3\%$, therefore, $\rho_e^{c,4}$ lies between the last
simulated density $76.9\%$ and $77.3\%$, i.e. $\rho_e^{c,4} = 77.1\% \pm
0.2\%$.
The fact that there exists a gap between the end point of the 
critical phase $\rho_e^{c,1} \approx 78.1\%$ (taken from \cite{cla94}) and the 
end point of the mixed phase $\rho_e^{c,4}$ leaves open 
the possibility of the existence of a third phase containing the
three-dimensional analogon of spiral waves (scroll waves). Due to the finite
size of our sample, however, we could only observe two-dimensional, 
flat fire fronts in that region.

The results for all simulated lattices and dimensions can be seen in
Tab.~\ref{table1}.
 
\section{Mapping Self-Organized Criticality onto Criticality}
\label{seceight}

The model we have investigated in this paper is an example of a system
which is far from equilibrium. It shows a wealth of interesting 
structures depending on the density of empty sites $\rho_e$. 
For large $\rho_e$, we find a region of vanishing fire density
which contains the critical behavior of the SOC forest-fire model. 
For lower density of empty sites we obtain the spiral waves of
the Bak et al. model. Finally, for even lower $\rho_e$,
we find a phase in which the trees show the tendency to form clusters
with the fire burning at their edges. The transition between the second and 
the third phase is of first order, whereas the transition between the 
first and the second phase is rather unconventional. We find hysteresis
and critical behavior on only one side of the transition.
The close relation of the spiral wave phase with the synchronized phase of
the model of \cite{cla95}
allows us to understand the mechanism which determines the density
distribution in the spiral state of not only this model, but of
arbitrary excitable systems. In particular, it yields the
density in front of the excitation fronts, the minimum density of excitable 
constituents that is necessary to sustain the excitation, and the factor of 
proportionality between the distance of the spiral arms
and the 'tree growth' rate $p$. 

Apart from being interesting in its own right, the model presented in this 
paper (like the model of \cite{cla95}) has 
to be seen also in the context of the general 
claim of \cite{sor95}, that the critical points of SOC models can be regarded
as ordinary
critical points of second order phase transitions. It was claimed in 
\cite{sor95} that this should be possible for all SOC models. There were
also given instructions on how to achieve this goal
for some particular models including the forest-fire model. 

The results in this paper and in \cite{cla95}, however, disprove this
hypothesis.
While the sub-critical side of the transition in both models
behaves as expected, the other
side (the side with a supposed non-vanishing order parameter) 
always exhibits surprising features. 
In the model of \cite{cla95} we found a whole critical region and lots of 
first-order phase transitions, whereas the actual model shows hysteresis and no
criticality at all.

Although these 'negative' results can not 
strictly rule out the possibility that
for some further slight change of the rules one might succeed in obtaining
the usual 'decent' behavior of the order parameter, the argument at the end
of Sec.~\ref{secfour}, 
together with the simulation results of this paper and of
\cite{cla95} make it seem very unlikely. 
We consider it to be more probable that a subset of SOC models
are 'only' ordinary critical models 'in disguise'. With these models 
the mapping proposed in \cite{sor95} should be possible 
(the results in \cite{bag96} also seem
to point in that direction), while there also exist genuine
SOC models for which this procedure can not be carried through.  
The forest-fire model seems to belong to the second class.

We suggest that similar phenomena and difficulties (critical regions, 
hysteresis at the critical
point, critical behavior on only one side) will also be found
in other models of SOC, as e.g. the sandpile, earthquake, and evolution
models. 

This work was supported by the Deutsche 
Forschungsgemeinschaft (DFG) under Contract
No Schw 348/7-1.

We thank B. Drossel for fruitful discussions.

\section{Appendix: Some Properties of the Order Parameter Curve}

All order parameter curves considered in the previous sections 
started at the point $\rho_f = \rho_t = 1/2, \rho_e = 0$. This starting
point is independent of dimension or lattice type, which can be seen as 
follows. For $\rho_e = 0$ each tree will sooner or later be set on fire by 
one of its $z$ nearest neighbors. 
Once having set on fire the tree, the fire jumps forever
between this site and its neighbors, because its neighbors,
after burning down, have to become trees again in the next time step
(there are no empty sites). The dynamics of these $1 + z$ sites 
then is fixed. After each time step trees and fires change places. But
the fire can still propagate to other sites, so that finally the whole system
consists of such 'blinking' regions. The state is periodic with period 2 and
the average fire and tree densities have to be $1/2$. For a random initial
state the densities are $1/2$ also without averaging.

The slope of $\rho_f(\rho_e)$ for $\rho_e \to 0$ depends only on the number 
of nearest neighbors $z$ and lies always within the interval $[-1;-1/2]$.
This can be seen as follows. If one takes a stationary state at $\rho_e = 0$ 
and inserts some empty sites by removing an equal number of trees and fires, 
the system will not remain in this state (with a slope of $-1/2$), 
but will adjust itself to a new stationary state with an even lower number
of fires, since the spreading conditions for the fire are now worse than before
due to the inserted empty sites. The magnitude of this effect depends only
on the coordination number $z$. The larger $z$, the smaller is the effect.
On an infinite-dimensional lattice,
the fire density will not readjust itself at all, because no fire can feel the 
empty sites. The slope in this case is the maximum possible slope $-1/2$.  
This can also be seen mathematically.
All fires have been trees in the preceding 
time step, therefore $\rho_f \le \rho_t$ in the stationary state. 
With $\rho_f + \rho_e + \rho_t = 1$ it
follows $\rho_f \le 1/2 - \rho_e/2$. Since for $\rho_e = 0$ both sides of
the last inequality are identical, one can differentiate and finds 
$\partial \rho_f / \partial \rho_e |_{\rho_e = 0} \le -1/2$. $1/2$ is the upper
bound for the slope.   

\begin{figure}
\psfig{file=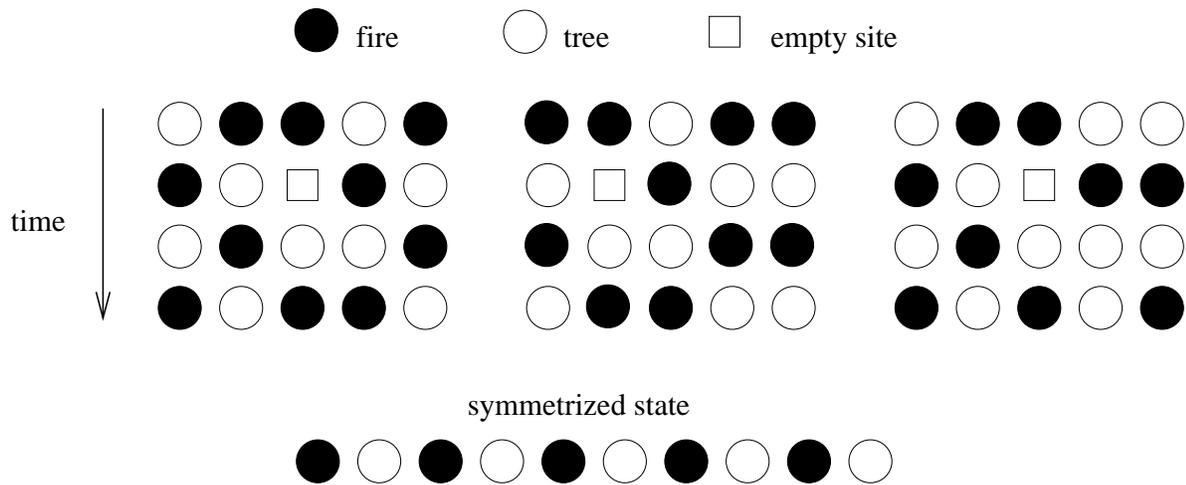,height=2.5in,angle=-90}
\vskip 1cm
\caption{The possible effects of the introduction of
one empty site into a one-dimensional system at $\rho_e = 0$.} 
\label{symm}
\end{figure}

The lower bound for the slope equals $-1$. This value is assumed 
for one dimension, as will be shown in the following
by considering the effect of introducing one empty site into the 
stationary 1D state at $\rho_e = 0$. Since the one-dimensional
lattice has the smallest possible number of neighbors, its slope represents
the lower bound. 
As argued in Sec.~\ref{secseven}, in the one-dimensional stationary state 
at $\rho_e = 0$
no more than two neighboring sites can be in the same state. 
In a symmetric state, where both neighbors of a tree are fires
and vice versa, i.e. where fires and trees are strictly alternating,
an empty site does not modify its neighborhood. 
Since the effects of the empty site are only local, it is sufficient to 
consider only three cases: One pair of sites in the same state (see left part 
of Fig.~\ref{symm}), collision of two pairs (see middle part of 
Fig.~\ref{symm}), and annihilation of two pairs 
(see right part of Fig.~\ref{symm}). The possibilities of placing the 
empty site in these cases can be further reduced to the cases shown
in the second line of Fig.~\ref{symm}. The reason is that, 
since in each time step $n$ trees burn down and are refilled 
into $n + 1$ empty sites, with probability $\approx 1$ for large $n$
the empty site at time $t$ was a fire at time $t-1$, i.e. at time $t$
the empty site occupies a place which would have been a tree in the 
unperturbed state.
Therefore, we place the empty site at a tree site in the second line of
Fig.~\ref{symm}. If instead we had chosen the neighboring tree sites, we would
only have generated mirror-symmetric forms of the configurations shown.
In the third line the empty site has disappeared from our
small section of the system and has become a tree with probability 
$\approx 1$ for $n \gg 1$. In the left case, the pair of sites
moves one lattice spacing due to the presence of the empty site. 
In the middle case, two pairs which are separated by only
one lattice site collide due to an empty site. 
In the right case, an empty site causes two neighboring pairs (which may have
collided earlier) to annihilate each other and thereby symmetrize the region 
around the empty site. 
Therefore, sooner or later all pairs will have annihilated, 
and the entire system (with the exception of the single empty site)
is in the symmetrical state. 
This evolution to the symmetrized state was also
observed in the simulations.

We can therefore restrict ourselves to consider
the consequences of the introduction of empty sites into such a highly
symmetrical state. This has the advantage that there is no need to consider the
neighboring sites, because they are not affected.
There are two possibilities to introduce
an empty site. The first one
is to remove a tree. In the next time step this tree would have
become a fire which it cannot do now. Instead, with very high probability
a new tree is grown at this empty site. During the re-growth of trees the empty
site effectively moves to another site in the system (which would have
been a 'tree site'). While there is one excess tree in one part of the system,
there is one tree 
less in another part of the system, so the total number of trees
does not change. The number of fires, however, has decreased by one, since
we prevented the 'birth' of one fire. The second possibility is to remove
a fire initially. If that happens, the empty site will have moved after the
tree-growth phase to a 'tree site', and we have the first case again.
Therefore, in each case
the introduction of empty sites takes place completely at the expense of fire 
sites and $\partial \rho_f / \partial \rho_e |_{\rho_e = 0} = -1$.

The measured results for the slope of the order parameter curve are shown
in Tab.~\ref{table1}. One can see that the slopes of the two-dimensional 
triangular
lattice and the three-dimensional hyper-cubic lattice are the same, since they 
have the same coordination number $z = 6$. 

\begin{figure}
\psfig{file=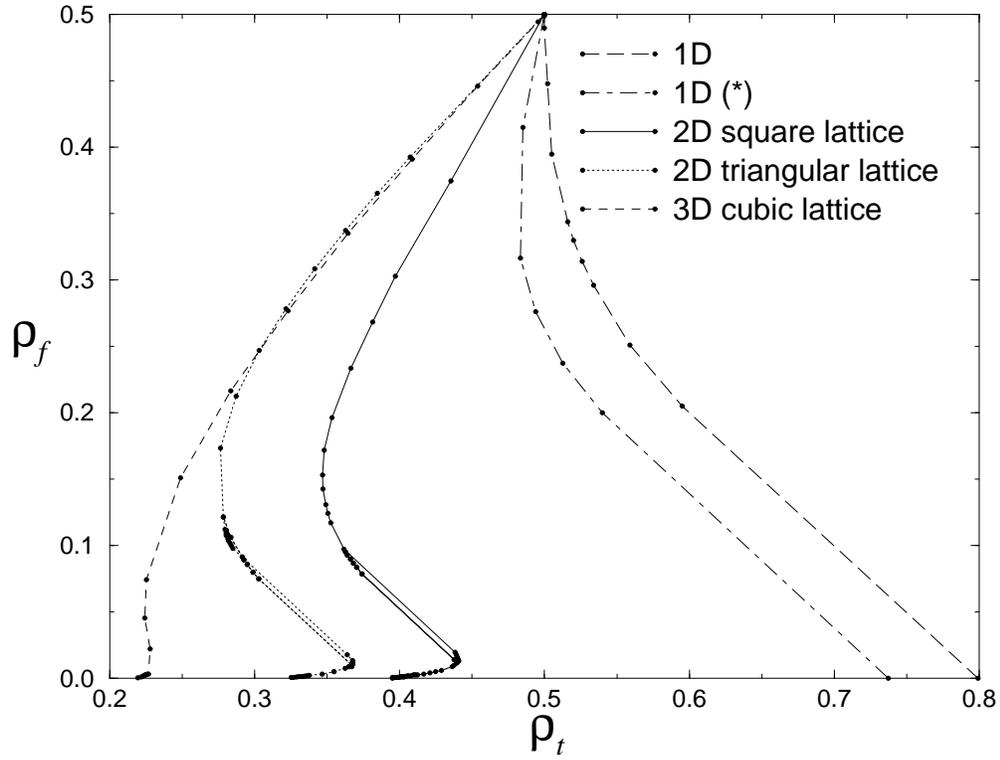,height=4in,angle=-90}
\vskip 1cm
\caption{The order parameter fire density $\rho_f$ for all simulated 
dimensions and lattice types as function of the tree density $\rho_t$. 
In the
case 1D (*) the fire was allowed to jump over one empty site, if necessary.}
\label{alldimensions}
\end{figure}

As mentioned earlier, not all tree densities between 0 and 1 are possible
in the stationary state. If one transforms the $\rho_f(\rho_e)$ curves to 
$\rho_f(\rho_t)$ curves via $\rho_f + \rho_t + \rho_e = 1$, one can
read off the diagram the allowed $\rho_t$ values 
(see Fig.~\ref{alldimensions}). 
For these values it is possible to simulate the model with fixed $\rho_t$.
The results do not differ from the results with fixed $\rho_e$.
Also in this case, one
has to be careful with the initial conditions, since there are sometimes  
two stationary 
states for the same value of $\rho_t$, and the initial state determines
which of the two possible stationary states will be chosen by the system.

\begin{table}
\begin{tabular}{llllll}
lattice type & 1D & 1D (*) & 2D & 2D triangular & 3D \\
\tableline
$\rho_e^{c,1}$ & $0\%$   & $0\%$   &$59.2(1)\%$&$66.4(4)\%^1$& $78.1(1)\%^1$ \\
$\rho_e^{c,2}$ & -       & -       & $60.8(5)\%$ & $67.7(5)\%$ & - \\
$\rho_e^{c,3}$ & -       & -       & $54.2\%$    & $61.7\%$    & - \\
$\rho_e^{c,4}$ & $20\%$  & $26\%$  & $54.7\%$    & $62.2\%$    & $77.1(2)\%$ \\
$-\frac{\left.\partial \rho_f}{\partial \rho_e\right|_{\rho_e = 0}}$    
               & 1.00(1) & 0.85(1) & 0.67(1)     & 0.54(1)     & 0.55(1) \\
\end{tabular}
\caption{The simulation results for various dimensions and lattice types
(${}^1$ = taken from [8]). 
In the case 1D (*) the fire was allowed to jump over one empty site, 
if necessary.}
\label{table1}
\end{table}
%
%

\end{document}

%% file: psfig_19.tex
\def\PsfigVersion{1.9}
\ifx\undefined\psfig\else \fi

%

\let\LaTeXAtSign=\@
\let\@=\relax
\edef\psfigRestoreAt{\catcode`\@=\number\catcode`@\relax}
\catcode`\@=11\relax
\newwrite\@unused
\def\ps@typeout#1{{\let\protect\string\immediate\write\@unused{#1}}}
\ps@typeout{psfig/tex \PsfigVersion}


\def\figurepath{./}
\def\psfigurepath#1{\edef\figurepath{#1}}

%
%
\def\@nnil{\@nil}
\def\@empty{}
\def\@psdonoop#1\@@#2#3{}
\def\@psdo#1:=#2\do#3{\edef\@psdotmp{#2}\ifx\@psdotmp\@empty \else
    \expandafter\@psdoloop#2,\@nil,\@nil\@@#1{#3}\fi}
\def\@psdoloop#1,#2,#3\@@#4#5{\def#4{#1}\ifx #4\@nnil \else
       #5\def#4{#2}\ifx #4\@nnil \else#5\@ipsdoloop #3\@@#4{#5}\fi\fi}
\def\@ipsdoloop#1,#2\@@#3#4{\def#3{#1}\ifx #3\@nnil 
       \let\@nextwhile=\@psdonoop \else
      #4\relax\let\@nextwhile=\@ipsdoloop\fi\@nextwhile#2\@@#3{#4}}
\def\@tpsdo#1:=#2\do#3{\xdef\@psdotmp{#2}\ifx\@psdotmp\@empty \else
    \@tpsdoloop#2\@nil\@nil\@@#1{#3}\fi}
\def\@tpsdoloop#1#2\@@#3#4{\def#3{#1}\ifx #3\@nnil 
       \let\@nextwhile=\@psdonoop \else
      #4\relax\let\@nextwhile=\@tpsdoloop\fi\@nextwhile#2\@@#3{#4}}
%
\ifx\undefined\fbox
\newdimen\fboxrule
\newdimen\fboxsep
\newdimen\ps@tempdima
\newbox\ps@tempboxa
\fboxsep = 3pt
\fboxrule = .4pt
\long\def\fbox#1{\leavevmode\setbox\ps@tempboxa\hbox{#1}\ps@tempdima\fboxrule
    \advance\ps@tempdima \fboxsep \advance\ps@tempdima \dp\ps@tempboxa
   \hbox{\lower \ps@tempdima\hbox
  {\vbox{\hrule height \fboxrule
          \hbox{\vrule width \fboxrule \hskip\fboxsep
          \vbox{\vskip\fboxsep \box\ps@tempboxa\vskip\fboxsep}\hskip 
                 \fboxsep\vrule width \fboxrule}
                 \hrule height \fboxrule}}}}
\fi
%
%
\newread\ps@stream
\newif\ifnot@eof       
\newif\if@noisy        
\newif\if@atend        
\newif\if@psfile       
%
%
{\catcode`\%=12\global\gdef\epsf@start{
\def\epsf@PS{PS}
\def\epsf@getbb#1{%
%
%
\openin\ps@stream=#1
\ifeof\ps@stream\ps@typeout{Error, File #1 not found}\else
%
%
   {\not@eoftrue \chardef\other=12
    \def\do##1{\catcode`##1=\other}\dospecials \catcode`\ =10
    \loop
       \if@psfile
	  \read\ps@stream to \epsf@fileline
       \else{
	  \obeyspaces
          \read\ps@stream to \epsf@tmp\global\let\epsf@fileline\epsf@tmp}
       \fi
       \ifeof\ps@stream\not@eoffalse\else
%
%
       \if@psfile\else
       \expandafter\epsf@test\epsf@fileline:. \\%
       \fi
%
%
          \expandafter\epsf@aux\epsf@fileline:. \\%
       \fi
   \ifnot@eof\repeat
   }\closein\ps@stream\fi}%
%
%
\long\def\epsf@test#1#2#3:#4\\{\def\epsf@testit{#1#2}
			\ifx\epsf@testit\epsf@start\else
\ps@typeout{Warning! File does not start with `\epsf@start'.  It may not be a PostScript file.}
			\fi
			\@psfiletrue} 
%
%
{\catcode`\%=12\global\let\epsf@percent=
%
%
%
\long\def\epsf@aux#1#2:#3\\{\ifx#1\epsf@percent
   \def\epsf@testit{#2}\ifx\epsf@testit\epsf@bblit
	\@atendfalse
        \epsf@atend #3 . \\%
	\if@atend	
	   \if@verbose{
		\ps@typeout{psfig: found `(atend)'; continuing search}
	   }\fi
        \else
        \epsf@grab #3 . . . \\%
        \not@eoffalse
        \global\no@bbfalse
        \fi
   \fi\fi}%
%
%
\def\epsf@grab #1 #2 #3 #4 #5\\{%
   \global\def\epsf@llx{#1}\ifx\epsf@llx\empty
      \epsf@grab #2 #3 #4 #5 .\\\else
   \global\def\epsf@lly{#2}%
   \global\def\epsf@urx{#3}\global\def\epsf@ury{#4}\fi}%
%
%
\def\epsf@atendlit{(atend)} 
\def\epsf@atend #1 #2 #3\\{%
   \def\epsf@tmp{#1}\ifx\epsf@tmp\empty
      \epsf@atend #2 #3 .\\\else
   \ifx\epsf@tmp\epsf@atendlit\@atendtrue\fi\fi}


\chardef\psletter = 11 
\chardef\other = 12

\newif \ifdebug 
\newif\ifc@mpute 
\c@mputetrue 

\let\then = \relax
\def\r@dian{pt }
\let\r@dians = \r@dian
\let\dimensionless@nit = \r@dian
\let\dimensionless@nits = \dimensionless@nit
\def\internal@nit{sp }
\let\internal@nits = \internal@nit
\newif\ifstillc@nverging
\def \Mess@ge #1{\ifdebug \then \message {#1} \fi}

{ 
	\catcode `\@ = \psletter
	\gdef \nodimen {\expandafter \n@dimen \the \dimen}
	\gdef \term #1 #2 #3%
	       {\edef \t@ {\the #1}
		\edef \t@@ {\expandafter \n@dimen \the #2\r@dian}%
		\t@rm {\t@} {\t@@} {#3}%
	       }
	\gdef \t@rm #1 #2 #3%
	       {{%
		\count 0 = 0
		\dimen 0 = 1 \dimensionless@nit
		\dimen 2 = #2\relax
		\Mess@ge {Calculating term #1 of \nodimen 2}%
		\loop
		\ifnum	\count 0 < #1
		\then	\advance \count 0 by 1
			\Mess@ge {Iteration \the \count 0 \space}%
			\Multiply \dimen 0 by {\dimen 2}%
			\Mess@ge {After multiplication, term = \nodimen 0}%
			\Divide \dimen 0 by {\count 0}%
			\Mess@ge {After division, term = \nodimen 0}%
		\repeat
		\Mess@ge {Final value for term #1 of 
				\nodimen 2 \space is \nodimen 0}%
		\xdef \Term {#3 = \nodimen 0 \r@dians}%
		\aftergroup \Term
	       }}
	\catcode `\p = \other
	\catcode `\t = \other
	\gdef \n@dimen #1pt{#1} 
}

\def \Divide #1by #2{\divide #1 by #2} 

\def \Multiply #1by #2
       {{
	\count 0 = #1\relax
	\count 2 = #2\relax
	\count 4 = 65536
	\Mess@ge {Before scaling, count 0 = \the \count 0 \space and
			count 2 = \the \count 2}%
	\ifnum	\count 0 > 32767 
	\then	\divide \count 0 by 4
		\divide \count 4 by 4
	\else	\ifnum	\count 0 < -32767
		\then	\divide \count 0 by 4
			\divide \count 4 by 4
		\else
		\fi
	\fi
	\ifnum	\count 2 > 32767 
	\then	\divide \count 2 by 4
		\divide \count 4 by 4
	\else	\ifnum	\count 2 < -32767
		\then	\divide \count 2 by 4
			\divide \count 4 by 4
		\else
		\fi
	\fi
	\multiply \count 0 by \count 2
	\divide \count 0 by \count 4
	\xdef \product {#1 = \the \count 0 \internal@nits}%
	\aftergroup \product
       }}

\def\r@duce{\ifdim\dimen0 > 90\r@dian \then   
		\multiply\dimen0 by -1
		\advance\dimen0 by 180\r@dian
		\r@duce
	    \else \ifdim\dimen0 < -90\r@dian \then  
		\advance\dimen0 by 360\r@dian
		\r@duce
		\fi
	    \fi}

\def\Sine#1%
       {{%
	\dimen 0 = #1 \r@dian
	\r@duce
	\ifdim\dimen0 = -90\r@dian \then
	   \dimen4 = -1\r@dian
	   \c@mputefalse
	\fi
	\ifdim\dimen0 = 90\r@dian \then
	   \dimen4 = 1\r@dian
	   \c@mputefalse
	\fi
	\ifdim\dimen0 = 0\r@dian \then
	   \dimen4 = 0\r@dian
	   \c@mputefalse
	\fi
	\ifc@mpute \then
		\divide\dimen0 by 180
		\dimen0=3.141592654\dimen0
		\dimen 2 = 3.1415926535897963\r@dian 
		\divide\dimen 2 by 2 
		\Mess@ge {Sin: calculating Sin of \nodimen 0}%
		\count 0 = 1 
		\dimen 2 = 1 \r@dian 
		\dimen 4 = 0 \r@dian 
		\loop
			\ifnum	\dimen 2 = 0 
			\then	\stillc@nvergingfalse 
			\else	\stillc@nvergingtrue
			\fi
			\ifstillc@nverging 
			\then	\term {\count 0} {\dimen 0} {\dimen 2}%
				\advance \count 0 by 2
				\count 2 = \count 0
				\divide \count 2 by 2
				\ifodd	\count 2 
				\then	\advance \dimen 4 by \dimen 2
				\else	\advance \dimen 4 by -\dimen 2
				\fi
		\repeat
	\fi		
			\xdef \sine {\nodimen 4}%
       }}

\def\Cosine#1{\ifx\sine\UnDefined\edef\Savesine{\relax}\else
		             \edef\Savesine{\sine}\fi
	{\dimen0=#1\r@dian\advance\dimen0 by 90\r@dian
	 \Sine{\nodimen 0}
	 \xdef\cosine{\sine}
	 \xdef\sine{\Savesine}}}	      

\def\psdraft{
	\def\@psdraft{0}
}
\def\psfull{
	\def\@psdraft{100}
}

\psfull

\newif\if@scalefirst
\def\psscalefirst{\@scalefirsttrue}
\def\psrotatefirst{\@scalefirstfalse}
\psrotatefirst

\newif\if@draftbox
\def\psnodraftbox{
	\@draftboxfalse
}
\def\psdraftbox{
	\@draftboxtrue
}
\@draftboxtrue

\newif\if@prologfile
\newif\if@postlogfile
\def\pssilent{
	\@noisyfalse
}
\def\psnoisy{
	\@noisytrue
}
\psnoisy
\newif\if@bbllx
\newif\if@bblly
\newif\if@bburx
\newif\if@bbury
\newif\if@height
\newif\if@width
\newif\if@rheight
\newif\if@rwidth
\newif\if@angle
\newif\if@clip
\newif\if@verbose
\def\@p@@sclip#1{\@cliptrue}

\newif\if@decmpr


\def\@p@@sfigure#1{\def\@p@sfile{null}\def\@p@sbbfile{null}
	        \openin1=#1.bb
		\ifeof1\closein1
	        	\openin1=\figurepath#1.bb
			\ifeof1\closein1
			        \openin1=#1
				\ifeof1\closein1%
				       \openin1=\figurepath#1
					\ifeof1
					   \ps@typeout{Error, File #1 not found}
						\if@bbllx\if@bblly
				   		\if@bburx\if@bbury
			      				\def\@p@sfile{#1}%
			      				\def\@p@sbbfile{#1}%
							\@decmprfalse
				  	   	\fi\fi\fi\fi
					\else\closein1
				    		\def\@p@sfile{\figurepath#1}%
				    		\def\@p@sbbfile{\figurepath#1}%
						\@decmprfalse
	                       		\fi%
			 	\else\closein1%
					\def\@p@sfile{#1}
					\def\@p@sbbfile{#1}
					\@decmprfalse
			 	\fi
			\else
				\def\@p@sfile{\figurepath#1}
				\def\@p@sbbfile{\figurepath#1.bb}
				\@decmprtrue
			\fi
		\else
			\def\@p@sfile{#1}
			\def\@p@sbbfile{#1.bb}
			\@decmprtrue
		\fi}

\def\@p@@sfile#1{\@p@@sfigure{#1}}

\def\@p@@sbbllx#1{
		\@bbllxtrue
		\dimen100=#1
		\edef\@p@sbbllx{\number\dimen100}
}
\def\@p@@sbblly#1{
		\@bbllytrue
		\dimen100=#1
		\edef\@p@sbblly{\number\dimen100}
}
\def\@p@@sbburx#1{
		\@bburxtrue
		\dimen100=#1
		\edef\@p@sbburx{\number\dimen100}
}
\def\@p@@sbbury#1{
		\@bburytrue
		\dimen100=#1
		\edef\@p@sbbury{\number\dimen100}
}
\def\@p@@sheight#1{
		\@heighttrue
		\dimen100=#1
   		\edef\@p@sheight{\number\dimen100}
}
\def\@p@@swidth#1{
		\@widthtrue
		\dimen100=#1
		\edef\@p@swidth{\number\dimen100}
}
\def\@p@@srheight#1{
		\@rheighttrue
		\dimen100=#1
		\edef\@p@srheight{\number\dimen100}
}
\def\@p@@srwidth#1{
		\@rwidthtrue
		\dimen100=#1
		\edef\@p@srwidth{\number\dimen100}
}
\def\@p@@sangle#1{
		\@angletrue
		\edef\@p@sangle{#1} 
}
\def\@p@@ssilent#1{ 
		\@verbosefalse
}
\def\@p@@sprolog#1{\@prologfiletrue\def\@prologfileval{#1}}
\def\@p@@spostlog#1{\@postlogfiletrue\def\@postlogfileval{#1}}
\def\@cs@name#1{\csname #1\endcsname}
\def\@setparms#1=#2,{\@cs@name{@p@@s#1}{#2}}
%
%
\def\ps@init@parms{
		\@bbllxfalse \@bbllyfalse
		\@bburxfalse \@bburyfalse
		\@heightfalse \@widthfalse
		\@rheightfalse \@rwidthfalse
		\def\@p@sbbllx{}\def\@p@sbblly{}
		\def\@p@sbburx{}\def\@p@sbbury{}
		\def\@p@sheight{}\def\@p@swidth{}
		\def\@p@srheight{}\def\@p@srwidth{}
		\def\@p@sangle{0}
		\def\@p@sfile{} \def\@p@sbbfile{}
		\def\@p@scost{10}
		\def\@sc{}
		\@prologfilefalse
		\@postlogfilefalse
		\@clipfalse
		\if@noisy
			\@verbosetrue
		\else
			\@verbosefalse
		\fi
}
%
%
\def\parse@ps@parms#1{
	 	\@psdo\@psfiga:=#1\do
		   {\expandafter\@setparms\@psfiga,}}
%
%
\newif\ifno@bb
\def\bb@missing{
	\if@verbose{
		\ps@typeout{psfig: searching \@p@sbbfile \space  for bounding box}
	}\fi
	\no@bbtrue
	\epsf@getbb{\@p@sbbfile}
        \ifno@bb \else \bb@cull\epsf@llx\epsf@lly\epsf@urx\epsf@ury\fi
}	
\def\bb@cull#1#2#3#4{
	\dimen100=#1 bp\edef\@p@sbbllx{\number\dimen100}
	\dimen100=#2 bp\edef\@p@sbblly{\number\dimen100}
	\dimen100=#3 bp\edef\@p@sbburx{\number\dimen100}
	\dimen100=#4 bp\edef\@p@sbbury{\number\dimen100}
	\no@bbfalse
}
\newdimen\p@intvaluex
\newdimen\p@intvaluey
\def\rotate@#1#2{{\dimen0=#1 sp\dimen1=#2 sp
		  \global\p@intvaluex=\cosine\dimen0
		  \dimen3=\sine\dimen1
		  \global\advance\p@intvaluex by -\dimen3
		  \global\p@intvaluey=\sine\dimen0
		  \dimen3=\cosine\dimen1
		  \global\advance\p@intvaluey by \dimen3
		  }}
\def\compute@bb{
		\no@bbfalse
		\if@bbllx \else \no@bbtrue \fi
		\if@bblly \else \no@bbtrue \fi
		\if@bburx \else \no@bbtrue \fi
		\if@bbury \else \no@bbtrue \fi
		\ifno@bb \bb@missing \fi
		\ifno@bb \ps@typeout{FATAL ERROR: no bb supplied or found}
			\no-bb-error
		\fi
		%
%
		\count203=\@p@sbburx
		\count204=\@p@sbbury
		\advance\count203 by -\@p@sbbllx
		\advance\count204 by -\@p@sbblly
		\edef\ps@bbw{\number\count203}
		\edef\ps@bbh{\number\count204}
		\if@angle 
			\Sine{\@p@sangle}\Cosine{\@p@sangle}
	        	{\dimen100=\maxdimen\xdef\r@p@sbbllx{\number\dimen100}
					    \xdef\r@p@sbblly{\number\dimen100}
			                    \xdef\r@p@sbburx{-\number\dimen100}
					    \xdef\r@p@sbbury{-\number\dimen100}}
%
                        \def\minmaxtest{
			   \ifnum\number\p@intvaluex<\r@p@sbbllx
			      \xdef\r@p@sbbllx{\number\p@intvaluex}\fi
			   \ifnum\number\p@intvaluex>\r@p@sbburx
			      \xdef\r@p@sbburx{\number\p@intvaluex}\fi
			   \ifnum\number\p@intvaluey<\r@p@sbblly
			      \xdef\r@p@sbblly{\number\p@intvaluey}\fi
			   \ifnum\number\p@intvaluey>\r@p@sbbury
			      \xdef\r@p@sbbury{\number\p@intvaluey}\fi
			   }
			\rotate@{\@p@sbbllx}{\@p@sbblly}
			\minmaxtest
			\rotate@{\@p@sbbllx}{\@p@sbbury}
			\minmaxtest
			\rotate@{\@p@sbburx}{\@p@sbblly}
			\minmaxtest
			\rotate@{\@p@sbburx}{\@p@sbbury}
			\minmaxtest
			\edef\@p@sbbllx{\r@p@sbbllx}\edef\@p@sbblly{\r@p@sbblly}
			\edef\@p@sbburx{\r@p@sbburx}\edef\@p@sbbury{\r@p@sbbury}
		\fi
		\count203=\@p@sbburx
		\count204=\@p@sbbury
		\advance\count203 by -\@p@sbbllx
		\advance\count204 by -\@p@sbblly
		\edef\@bbw{\number\count203}
		\edef\@bbh{\number\count204}
}
%
%
\def\in@hundreds#1#2#3{\count240=#2 \count241=#3
		     \count100=\count240	
		     \divide\count100 by \count241
		     \count101=\count100
		     \multiply\count101 by \count241
		     \advance\count240 by -\count101
		     \multiply\count240 by 10
		     \count101=\count240	
		     \divide\count101 by \count241
		     \count102=\count101
		     \multiply\count102 by \count241
		     \advance\count240 by -\count102
		     \multiply\count240 by 10
		     \count102=\count240	
		     \divide\count102 by \count241
		     \count200=#1\count205=0
		     \count201=\count200
			\multiply\count201 by \count100
		 	\advance\count205 by \count201
		     \count201=\count200
			\divide\count201 by 10
			\multiply\count201 by \count101
			\advance\count205 by \count201
		     \count201=\count200
			\divide\count201 by 100
			\multiply\count201 by \count102
			\advance\count205 by \count201
		     \edef\@result{\number\count205}
}
\def\compute@wfromh{
		\in@hundreds{\@p@sheight}{\@bbw}{\@bbh}
		\edef\@p@swidth{\@result}
}
\def\compute@hfromw{
	        \in@hundreds{\@p@swidth}{\@bbh}{\@bbw}
		\edef\@p@sheight{\@result}
}
\def\compute@handw{
		\if@height 
			\if@width
			\else
				\compute@wfromh
			\fi
		\else 
			\if@width
				\compute@hfromw
			\else
				\edef\@p@sheight{\@bbh}
				\edef\@p@swidth{\@bbw}
			\fi
		\fi
}
\def\compute@resv{
		\if@rheight \else \edef\@p@srheight{\@p@sheight} \fi
		\if@rwidth \else \edef\@p@srwidth{\@p@swidth} \fi
}
%
\def\compute@sizes{
	\compute@bb
	\if@scalefirst\if@angle
	\if@width
	   \in@hundreds{\@p@swidth}{\@bbw}{\ps@bbw}
	   \edef\@p@swidth{\@result}
	\fi
	\if@height
	   \in@hundreds{\@p@sheight}{\@bbh}{\ps@bbh}
	   \edef\@p@sheight{\@result}
	\fi
	\fi\fi
	\compute@handw
	\compute@resv}

%
%
\def\psfig#1{\vbox {
	%
	\ps@init@parms
	\parse@ps@parms{#1}
	\compute@sizes
	\ifnum\@p@scost<\@psdraft{
		\special{ps::[begin] 	\@p@swidth \space \@p@sheight \space
				\@p@sbbllx \space \@p@sbblly \space
				\@p@sbburx \space \@p@sbbury \space
				startTexFig \space }
		\if@angle
			\special {ps:: \@p@sangle \space rotate \space} 
		\fi
		\if@clip{
			\if@verbose{
				\ps@typeout{(clip)}
			}\fi
			\special{ps:: doclip \space }
		}\fi
		\if@prologfile
		    \special{ps: plotfile \@prologfileval \space } \fi
		\if@decmpr{
			\if@verbose{
				\ps@typeout{psfig: including \@p@sfile.Z \space }
			}\fi
			\special{ps: plotfile "`zcat \@p@sfile.Z" \space }
		}\else{
			\if@verbose{
				\ps@typeout{psfig: including \@p@sfile \space }
			}\fi
			\special{ps: plotfile \@p@sfile \space }
		}\fi
		\if@postlogfile
		    \special{ps: plotfile \@postlogfileval \space } \fi
		\special{ps::[end] endTexFig \space }
		\vbox to \@p@srheight sp{
			\hbox to \@p@srwidth sp{
				\hss
			}
		\vss
		}
	}\else{
		\if@draftbox{		
			\hbox{\frame{\vbox to \@p@srheight sp{
			\vss
			\hbox to \@p@srwidth sp{ \hss \@p@sfile \hss }
			\vss
			}}}
		}\else{
			\vbox to \@p@srheight sp{
			\vss
			\hbox to \@p@srwidth sp{\hss}
			\vss
			}
		}\fi

	}\fi
}}
\psfigRestoreAt
\let\@=\LaTeXAtSign